\documentclass[preprint,aip]{revtex4}

\usepackage{graphicx}
\usepackage{dcolumn}
\usepackage{bm}

\begin{document}
\title{Non-magnetic simplified cylindrical cloak with suppressed $zero^{th}$ order scattering}
\date{\today}
\author{Wei Yan, Min Yan}
\author{Min Qiu}
\email{min@kth.se} \affiliation{Laboratory of Optics, Photonics and
Quantum Electronics, Department of Microelectronics and Applied
Physics, Royal Institute of Technology (KTH), Electrum 229, 16440
Kista, Sweden}
\date{\today}
\begin{abstract}
A type of simplified cloaks with matched exterior boundaries is
proposed. The cloak uses non-magnetic material for the TM
polarization and can function with a relatively thin thickness. It
is shown that the $zero^{th}$ order scattering of such cloak is
dominant among all cylindrical scattering terms. A gap is added at
the cloak's inner surface to eliminate the $zero^{th}$ order
scattering, through the mechanism of scattering resonance. The
reduction in scattering is relatively smooth, indicating that the
proposed scattering reduction method has good tolerance to
perturbations. Numerical simulations also confirm that the proposed
structure has very low scattering.

\end{abstract}
\maketitle

Recently, invisibility cloaks have attracted intense attentions due
to their amazing optical properties
\cite{P,L1,S,C1,Cm,R,C,L2,YM1,YM2,C2}. An ideal invisibility cloak
can exclude light from a protected region without perturbing the
exterior fields. The parameters of invisibility cloaks can be easily
obtained from coordinate transformation method \cite{P}. According
to the number of transformed coordinates, invisibility cloaks can be
classified into two-dimensional ones (line-transformed cloak) or
three-dimensional ones (point-transformed cloak) \cite{Yan2}. Most
of publications on invisibility cloaks are focused on
two-dimensional cloaks, especially cylindrical cloaks
\cite{S,C1,Cm,R,YM1,YM2,C2}.

As described in Ref. [1], a two-dimensional cylindrical cloak can be
easily constructed by compressing a cylindrical region $r^{'}<b$
into a concentric cylindrical shell $a<r<b$ radially. The coordinate
transformation is described by the function $r^{'}=f(r)$ with
$f(a)=0$ and $f(b)=b$. Under free space background, the relative
permittivity and permeability of cloak are as following
$\epsilon_r=\mu_r=f(r)/[rf^{'}(r)]$,
$\epsilon_\theta=\mu_\theta=rf^{'}(r)/f(r)$ and
$\epsilon_z=\mu_z=f(r)f^{'}(r)/r$, where $f^{'}(r)=df(r)/dr$.
Choosing various $f(r)$, we can obtain cloaks with different
parameters. However, $\epsilon_\theta$ and $\mu_\theta$ are always
infinitely large at the inner boundary $r=a$ regardless of $f(r)$
owing to $f(a)=0$. Thus, in order to avoid this unphysical
singularity and make the implementation practical, we should
simplify the parameters of cloak. The simplification principle is to
keep $\epsilon_i\mu_z$ and $\mu_i\epsilon_z$ ($i=r,\theta$) the same
as the ideal case \cite{S,Cm}.

In Ref. [3], the simplified linear cloak with
$\epsilon_z=[b/(b-a)]^2$, $\mu_\theta=1$ and $\mu_r=[(r-a)/r]^2$, is
proposed and experimentally demonstrated for the TE polarization.
However, as pointed in Ref. [9], this simplified cloak leads to
considerable scattering for both $zero^{th}$ order and high order
cylindrical waves. A major factor for the high scatterings is due to
mismatched exterior boundary. In Refs. [10] and [11], the authors
proposed two different types of simplified cloaks with matched
exterior boundary. The high order scatterings for such cloaks are
reduced substantially, due to matched exterior boundary. Hence, the
cloaking performance is improved significantly. However, as pointed
in Ref. [10], the $zero^{th}$ order scattering is still
considerable. It indicates that the cloak operates similarly as a
small-sized cylindrical rod, for which the scatterings are mainly
contributed from the $zero^{th}$ order scattering. As shown in Ref.
[12], the $zero^{th}$ order scattering for the small size rod can be
eliminated by employing an appropriate layer outside the rod. This
idea can also be utilized to the cloak structure. In this paper, we
will analyze how to eliminate the $zero^{th}$ order scattering of
simplified cloaks by employing an appropriate layer.

It is worth noting that the simplified cloaks proposed in Ref. [10]
have drawbacks of requiring magnetic materials even for a single
polarization, while the cloaked proposed in Ref [11] having the size
restriction of $b/a>2$. In order to overcome these drawbacks, here
we propose another transformation function
\begin{equation}
f(r)=\frac{-ar^2+(b^2+a^2)r-ab^2}{(b-a)^2}.
\end{equation}
The parameters of a cloak obtained from such a transformation are:
\begin{equation}
\epsilon _r = \mu_r=\frac{{(b^2  - ra)^2(r - a)^2}}{{(b - a)^4
r^2}},\;\varepsilon _\theta=\mu_\theta  = \frac{{(a^2  + b^2 -
2ra)^2}}{{(b - a)^4 }},\;\epsilon_z=\mu_z=1.
\end{equation}
It is seen from Eq. (2) that the parameters of the cloak at the
exterior boundary is the same as free space, i.e.,  the exterior
boundary matches perfectly with free space. Since $\mu_z=1$, the
cloak can be made non-magnetic for the TM polarization. In
particular, different from Ref. [11], the cloak proposed in the
present paper does not have any size restrictions, since $f(r)$ is
always a monotonic increasing function at $a<r<b$. The monotonic
increasing property is clearly seen from $f^{'}(r)=(a^2 + b^2 -
2ra)/(b - a)^2$, which is always positive at $a<r<b$.

In the following, our discussions will be focused on the cloak with
parameters expressed in Eq. (2). Only the TM polarization will be
considered, since the cloak can be made non-magnetic under this
polarization. In order to make the performance of the cloak
independent of the material inside the cloaked region, a PEC lining
is put at the boundary of the cloaked region [10]. To maintain the
matched exterior boundary, the embedded layer to eliminate the
$zero^{th}$ order scattering is put at the inner boundary of the
cloak. Consider the simplest case that the embedded layer is just
free space, i.e., a gap is imposed at the cloak's inner surface. In
the following, we will focus on this case with the structure
illustrated in Fig. 1, where it is seen that a gap with a width $d$
is added between the cloak and the cloaked region.

To choose the appropriate thickness of the imposed gap, the
$zero^{th}$ order scattering coefficient of the proposed structure
needs to be analyzed. Since $\epsilon_\theta$ varies only with $r$,
an asymptotic approach can be employed to calculate the $zero^{th}$
order scattering coefficient. We firstly divide the cloak into $N$
gradient layers. Then, we are able to calculate the $zero^{th}$
scattering coefficient or $R_N^{0sc}$, analytically using a matrix
method. $R_N^{0sc}$ is characterized by
\begin{equation}
DC_N  \cdots C_1 C_0 \left[ {\begin{array}{*{20}c}
   1  \\
   {R_N ^{0sc} }  \\
\end{array}} \right] =0,
\end{equation}
with
\begin{equation}
C_n  = \left[ {\begin{array}{*{20}c}
   {J_0 (k_{n{\rm{ + 1}}} r_{n} )} & {H_0 (k_{n{\rm{ + 1}}} r_{n{}} )}  \\
   {\eta _{n + 1} J_0 ^{\rm{'}} (k_{n{}} r_{n{}} )} & {\eta _{n + 1} H_0 ^{\rm{'}} (k_{n{\rm{ + 1}}} r_{n{\rm{ + 1}}} )}  \\
\end{array}} \right]^{ - 1}\left[ {\begin{array}{*{20}c}
   {J_0 (k_n r_n )} & {H_0 (k_n r_n )}  \\
   {\eta _n J_0 ^{'} (k_n r_n )} & {\eta _n H_0 ^{'} (k_n r_n )}  \\
\end{array}} \right],
\end{equation}
\begin{equation}
D = \left[ {\begin{array}{*{20}c}
   {\eta _{0} J_0 ^{\rm{'}} (k_{{0}} r_{{\rm{N + 1}}} )} & {\eta _{0} H_0 ^{\rm{'}} (k_{{\rm{0}}} r_{{\rm{N + 1}}} )}  \\
\end{array}} \right],
\end{equation}
where $r_{N+1}=a-d$ and $r_n=b+(b-a)n/N$ ($n=0,1,\cdots N$);
$k_n=\omega\sqrt {\varepsilon _\theta(r_n) }/c$ with $\eta _n=\sqrt
{1/\varepsilon _\theta(r_n)}$ ($n=0,1,\cdots N$).

From Eq. (3), $R_N ^{0sc}$ as a function of $d$ can be easily
obtained. Finally, the value of $d$ for eliminating the $zero^{th}$
order scattering can be approached by solving $\mathop {\lim
}\limits_{N \to \infty }R_N ^{0sc}(d)=0$. For an example when
$a=0.3m$, $b=0.6m$, and the wavelength $\lambda$ is $0.3m$, the
value $d$ for $\mathop {\lim }\limits_{N \to \infty }R_N
^{0sc}(d)=0$ is calculated to be $0.1365m$.

To illustrate the effect of the added gap, we carry out simulations
for this example with commercial COMSOL Mutiphysics package. For
comparison, the case without the gap is also simulated. The
scattering coefficients versus wavelength for these two cases are
plotted in Fig. 2. It is seen that the $zero^{th}$ order scattering
reduces dramatically around $\lambda=0.3m$ by employing the gap,
which agrees well with above analysis. This cancellation phenomenon
is due to the scattering resonance. It is also seen from Fig. 2(b)
that the $zero^{th}$ order scattering approaches zero quite
smoothly. Within certain wavelength range near $\lambda=0.3m$, the
$zero^{th}$ order scattering is relatively small. This in turn means
that the $zero^{th}$ order scattering at $\lambda=0.3m$ should be
very small even when the thickness of the gap deviate from ideal
design values. Thus, this effect is relatively insensitive to
perturbations of the structure. Besides the dip in the scattering,
there exists a peak around $\lambda=0.23m$, which is also due to the
anti scattering resonance. For high order scattering terms, no
obvious difference between two cases is observed. It is due to the
fact that the fields are almost zero near the inner boundary of
cloak for high order terms [9].

In Fig. 3(a) and 3(b), snapshots of $H_z$ fields at $\lambda=0.3m$
for the case without the gap and the case with the gap are plotted,
respectively. It is clearly seen that the performance of the cloak
is improved by imposing the gap. The difference can be further
confirmed in Fig. 3(c) and 3(d), where only the scattering fields
for two cases are plotted. Considerable reduction in scattering is
observed. In Fig. (4), the radar cross section (RCS) normalized by
the wavelength is plotted for the bare PEC cylinder, the cloak
without the gap, and the cloak with the gap, respectively. It is
seen that the RCS is reduced dramatically by imposing such a gap.
The RCS of backscattering, i.e., at $\theta=180^o$, is about $5dB$
and $-0.5dB$ for the bare PEC cylinder and the cloak without the
gap; while the RCS of the backscattering is only $-8dB$ for the
cloak with the gap. RCS also oscillates with angles much faster,
which indicates the $zero^{th}$ order scattering is eliminated and
the high order scatterings dominate.

In conclusion, we proposed a type of simplified cloaks, which is
non-magnetic for the TM polarization and has no any restrictions on
cloak thickness. Since the designed simplified cloaks suffer from a
high $zero^{th}$ order scattering, we then proposed a method to
eliminate the $zero^{th}$ order scattering. This is achieved by
imposing a gap between the cloak and the cloaked region. Due to the
destructive scattering resonance, we can achieve complete
cancellation of the $zero^{th}$ order scattering for certain
thickness at the desired wavelength. The overall scattering of the
cloak is thus significantly reduced.

\newpage
\section*{Figure captions}
\textbf{Figure 1}: (Color online) Schematic picture of the proposed
cloak structure. A gap of free space with width $d$ is imposed
between the cloak and the PEC lining.

\textbf{Figure 2}: (Color online) Scattering coefficients of
cylindrical versus wavelength for the case (a) without the gap and
the case (b) with the gap. The structure parameters are $a=0.3m$,
$b=0.6m$.

\textbf{Figure 3}: (Color online) Snapshots of $H_z$ fields at
$\lambda=0.3m$ for the case (a) without the gap, and the case (b)
with the gap; The scattered $H_z$ fields at $\lambda=0.3m$ for the
case (c) without the gap, and the case (d) with the gap. The
structure parameters are the same as in Fig. 2.

\textbf{Figure 4}: (Color online) RCS normalized by wavelength for
the bare PEC cylinder, the cloak without the gap, and the cloak with
the gap. The structure parameters are the same as in Fig. 2, and
$\lambda=0.3m$.

\newpage
\begin{figure}[htbp]\centering
\includegraphics[width=7cm]{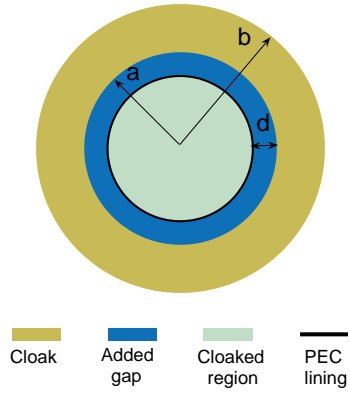}
\caption{(Color online) Schematic picture of the proposed cloak
structure. A gap of free space with width $d$ is imposed between the
cloak and the PEC lining.}
\end{figure}

\newpage
\begin{figure}[htbp]\centering
\includegraphics[width=6cm]{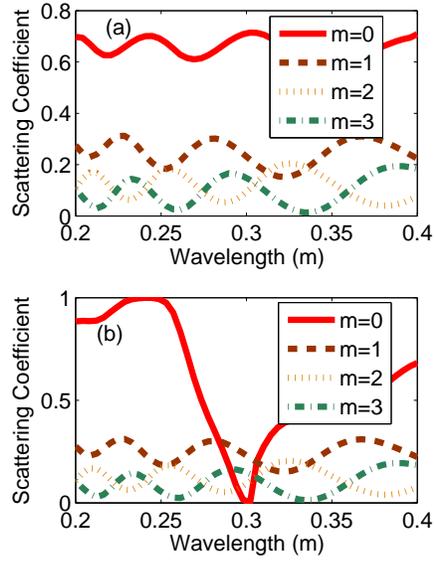}
\caption{(Color online) Scattering coefficients of cylindrical
versus wavelength for the case (a) without the gap and the case (b)
with the gap. The structure parameters are $a=0.3m$, $b=0.6m$.}
\end{figure}

\newpage
\begin{figure}[htbp]\centering
\includegraphics[width=8cm]{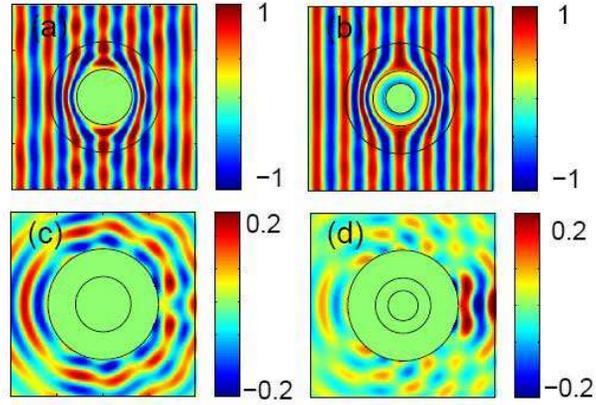}
\caption{(Color online) Snapshots of $H_z$ fields at $\lambda=0.3m$
for the case (a) without the gap, and the case (b) with the gap; The
scattered $H_z$ fields at $\lambda=0.3m$ for the case (c) without
the gap, and the case (d) with the gap. The structure parameters are
the same as in Fig. 2.}
\end{figure}
\newpage
\newpage
\begin{figure}[htbp]\centering
\includegraphics[width=7cm]{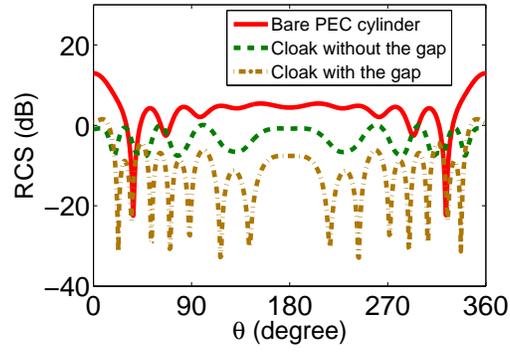}
\caption{(Color online) RCS normalized by wavelength for the bare
PEC cylinder, the cloak without the gap, and the cloak with the gap.
The structure parameters are the same as in Fig. 2, and
$\lambda=0.3m$.}
\end{figure}
\end{document}